\documentclass[sigconf]{acmart}
\usepackage{todonotes}
\usepackage{color}
\usepackage{soul} 

\usepackage[compact]{titlesec}
\titlespacing{\section}{1.1pt}{*0.5}{*0.5}
\titlespacing{\subsection}{0.1pt}{*0.1}{*0.1}
\titlespacing{\subsubsection}{0.1pt}{*0.1}{*0.1}

\usepackage{figsize}

\AtBeginDocument{%
  \providecommand\BibTeX{{%
    \normalfont B\kern-0.5em{\scshape i\kern-0.25em b}\kern-0.8em\TeX}}}

\setcopyright{iw3c2w3}
\copyrightyear{2021}
\acmYear{2021}
\acmDOI{10.1145/1122445.1122456}

\copyrightyear{2021}
\acmYear{2021}
\setcopyright{iw3c2w3}
\acmConference[WWW '21 Companion]{Companion Proceedings of the Web Conference 2021}{April 19--23, 2021}{Ljubljana, Slovenia}
\acmBooktitle{Companion Proceedings of the Web Conference 2021 (WWW '21 Companion), April 19--23, 2021, Ljubljana, Slovenia}
\acmPrice{}
\acmDOI{10.1145/3442442.3452327}
\acmISBN{978-1-4503-8313-4/21/04}

\begin{document}

\newcommand{\Tabledhote}{Table d'h\^{o}te}
\newcommand{\tabledhote}{table d'h\^{o}te}
\title{Toward the Next Generation of News Recommender Systems}

\author{Himan Abdollahpouri}
\affiliation{%
  \institution{Northwestern University}
  \city{Evanston}
  \state{Illinois}
  \country{USA}
  }
\email{himan.abdollahpouri@northwestern.com}

\author{Edward C.\ Malthouse}
\affiliation{%
  \institution{Northwestern University}
  \streetaddress{1845 Sheridan Road}
  \city{Evanston}
  \state{Illinois}
  \country{USA}
  \postcode{60645}}
\email{ecm@northwestern.com}

\author{Joseph A. Konstan}
\affiliation{%
  \institution{University of Minnesota}
  \city{Minneapolis}
  \state{Minnesota}
  \country{USA}
  }
\email{konstan@umn.edu}
\author{Bamshad Mobasher}
\affiliation{%
  \institution{DePaul University}
  \city{Chicago}
  \state{Illinois}
  \country{USA}
  }
\email{mobasher@cs.depaul.com}

\author{Jeremy Gilbert }
\affiliation{%
  \institution{Northwestern University}
  \city{Evanston}
  \state{Illinois}
  \country{USA}
  }
\email{jeremy.gilbert@northwestern.edu}

\renewcommand{\shortauthors}{Abdollahpouri, Malthouse, Konstan, Mobasher, Gilbert}

\begin{abstract}
This paper proposes a vision and research agenda for the next generation of news recommender systems (RS), called the \emph{\tabledhote\ approach}. A \tabledhote{} (translates as host's table) meal is a sequence of courses that create a balanced and enjoyable dining experience for a guest. Likewise, we believe news RS should strive to create a similar experience for the users by satisfying the news-diet needs of a user. While extant news RS considers criteria such as diversity and serendipity, and RS bundles have been studied for other contexts such as tourism, \tabledhote{} goes further by ensuring the recommended articles satisfy a diverse set of user needs in the right proportions and in a specific order. In \tabledhote{}, available articles need to be stratified based on the different ways that news can create value for the reader, building from theories and empirical research in journalism and user engagement. Using theories and empirical research from communication on the uses and gratifications (U\&G) consumers derive from media, we define two main strata in a \tabledhote{} news RS, each with its own substrata: 1) \emph{surveillance}, which consists of information the user needs to know, and 2) \emph{serendipity}, which are the articles offering unexpected surprises. The diversity of the articles according to the defined strata and the order of the articles within the list of recommendations are also two important aspects of the \tabledhote{} in order to give the users the most effective reading experience. We propose our vision, link it to the existing concepts in the RS literature, and identify challenges for future research. 
\end{abstract}





\maketitle
\section{Introduction}

There is over a quarter-century of research on news recommender systems (RS) \cite{lavrenko2000language,billsus1999personal, kamba1995krakatoa}. The majority of prior research, however, has focused on algorithms for estimating the user's utility of different news stories \cite[table~1]{karimi2018news}. There has been less work on how a list of news articles can be presented as a package. While there has been some work on creating diversity, novelty, and serendipity in news recommendations \cite{chakraborty2019optimizing}, we extend this line of research by offering a vision that builds on user-research from journalism and engagement that uncovers how news media contribute to personal goals in the user's life. By starting with user needs we can assemble a more balanced package of recommended articles.


Our vision is meant to apply to news RS in general, but we discuss the specific task of creating a newsletter in detail to motivate why our approach is needed. While newsletters can have a variety of goals, many are designed to quickly update the reader so as to avoid information overload. The goal is not to simply recommend items of interest to the user, but rather to create a self-contained package. The reader should be able to derive substantial value from reading the newsletter alone and feel updated possibly without clicking on any hyperlinks to read the entire stories. Since the median number of times a subscriber visits a news website is once per month \cite{mitchell2011}, newsletters and other outbound touchpoints such as e-editions are an important way for news organizations to offer value and retain paying subscribers. 

We see the task of creating, for example, a newsletter as one requiring careful curation, akin to an art exhibit or a fine meal, and call it the \textit{\tabledhote{} approach}, which means the host's table. A chef has composed a sequence of items that together form a balanced meal and create a dining experience, as opposed to offering an \`{a} la carte list of items that the user may like, or offering a buffet with hundreds of items. What we have in mind goes beyond simple diversification. While a basic diversified menu would offer dishes that are different from each other, a \tabledhote{} experience might offer surprise courses that are sequenced and complement each other. One course may set up the next, and the meal has been scripted to take the guest on a journey in the same way a symphony takes the listener through a series of moods and emotions, where a slow, quiet passage may set up a lively, intense one. Our vision for news RS shares these goals.

While a good editor could create what we have in mind, our goal is to create personalized newsletters for individual users. The \tabledhote{} approach has five main considerations: (1) the selected articles must cover what the user needs to know (surveillance), (2) there should also be unexpected, surprise articles mixed in (news serendipity), (3) there should be diversity and balance among the articles selected, (4) the order and sequencing in which the articles are presented matters, and (5) the articles should be carefully curated to avoid information overload. To create such newsletters in an automated way, the \tabledhote{} approach first stratifies the articles by how they satisfy a user need and then selects articles from different strata to achieve the multiple goals we have identified.

\section{Diversity and Balance in RS}

Research in RS started by trying to predict the rating a user might give to an item to estimate how relevant it could be. As a result, RS algorithms were mainly evaluated on how accurate their predictions were and whether the recommended items were relevant to the users' taste. However, researchers later realized this overemphasis on accuracy could actually hurt the success of an RS \cite{mcnee2006being}. Factors such as the diversity of the items within a list (i.e., the dissimilarity between the items in a list) became important for a successful RS \cite{kaminskas2016diversity}. The \tabledhote\ approach differs from the recommendation diversity literature as a diverse list according to news topics may not necessarily cover important user needs. A list of articles that contains a sports article, a business article, and an article about some celebrities is quite diverse but if there are some important articles such as a curfew lockdown that are not in the list, the list is lacking important information. 

The task of assembling a set of items as a package has also received some attention \cite{zhu2014bundle,bai2019personalized}. For example, in RS for tourism, a recommendation set is often built as a \textit{bundle} where the set can cover different needs of a user for a particular trip such as the hotel, flight, meals, and the rental car, all as one list of recommendations. In other words, package recommendation is to create a diverse set where not only the items within the package are dissimilar, but they also complement each other for a certain purpose. 

Although the concept of diversity is well-studied for news \cite{tintarev2017presenting,chakraborty2019optimizing}, not much work has been done on package recommendations in the news domain. The \tabledhote{} view in RS extends package recommendation so that articles are selected to cover certain aspects of the user's needs in terms of news. News is inherently different from many other domains because users do not read the news just to follow their interest (as opposed to watching movies or even reading a magazine) but to be also informed about what is happening in the world or in their community. The idea is, similar to how a chef may put together a set of items for a person that can form a balanced and diverse meal experience, the meal should also fulfill the person's need for necessary nutrition in order to stay healthy. In addition, the items within the package should often be consumed in a certain order to give the user the best experience. For instance, it is very rare for a meal to start with dessert, followed by the main course, and end with the appetizer.  


\section{\Tabledhote\ News RS}


The \tabledhote{} approach is intended to satisfy the news-diet needs of a user. A key \tabledhote{} assumption is that the items must achieve a certain level of diversity and balance. Diversity in the case of a great meal could be described by types of courses: one or more starter courses, building to the main course and closing with a finale course, perhaps with some surprise courses interspersed. Similarly, creating an automated \tabledhote{} newsletter will combine types (strata) of stories. The motivation comes from stratified sampling designs, where sampling units are first partitioned into strata and samples are then drawn from each one \cite{cochran1977sampling}.

We discuss how a \tabledhote{} news RS will classify the stories of the day into strata and select some number or proportion from each. The number/proportions from each stratum could be personalized to the individual user or it could be a fixed setup for every user based on the news company's mission and purpose. There are, however, many ways to define strata. Using theories and empirical research from communication on the uses and gratifications (U\&G) consumers derive from media \cite{Katz1973, Ruggiero2000}, we suggest two main strata, \textit{Surveillance} and \textit{Serendipity}, each with subdivisions. U\&G is conceptually similar to recent work on engagement by Lalmas et al.\ \citep{attfield2011towards, arapakis2014user,lagun2016understanding} that seeks to understand reader benefits from the news.

\subsection{Surveillance}
Perhaps the most fundamental need satisfied by news media is to inform, which is called the \emph{surveillance function of news}: ``Journalists fulfill people's innate desire to detect threats in the environment, keep informed about the world, and devise methods of dealing with these threats, whether real or potential'' \cite[p.\ 32]{shoemaker1996hardwired}. Such articles provide information that readers need to know in order to live their lives. U\&G theory discusses a related gratification labeled \emph{information}, which is defined as ``finding out about relevant events and conditions in immediate surroundings, society and the world; seeking advice on practical matters or opinion and decision choices; satisfying curiosity and general interest; learning, self-education; gaining a sense of security through knowledge'' \cite[p.\ 82]{Katz1973}.

We can further subdivide articles that fall into this surveillance stratum into those that \emph{all} readers need to know versus articles that \emph{some} need to know. This distinction maps directly to the RS algorithms. An article with information that everyone needs to know should be recommended to everyone and thus requires a non-personalized algorithm. Such stories include announcements (e.g., ``Curfew lockdown starts 10 pm tonight,''  the top picture in Figure \ref{fig:fig_examples}), emergency news (e.g., ``Hurricane forecast to strike in two days''), and important national or international news. We call these \emph{non-personalized surveillance articles} and expect them to be mostly human-curated.

There are also articles that some, but not all, readers need to know. Such articles lend themselves to personalized RS, and we will call them \emph{personalized surveillance articles}. A cancer patient needs to know about some new treatment, but this information is not essential to most readers. An investor who specializes in Asian securities and reads some financial newspaper may need to know about some development at the Tokyo stock exchange, but this information may not be essential for other investors. The middle picture in Figure \ref{fig:fig_examples} shows an example of this type of articles. 

There can be gradations of importance among such stories, which should be considered when ranking articles to recommend. For example, a reader might closely follow (surveil) the local football team and want to know every detail of the last game, and every rumor of a player trade. Having this knowledge may be a cornerstone of the person's identity, providing a way to connect with friends. A news organization may, nevertheless, want to prioritize personalized surveillance articles that concern the person's health or livelihood over a hobby like football. Algorithms ranking the articles may thus need to consider multiple objectives.

Personalized surveillance articles can be further stratified, which can facilitate diversification in recommendations. Surveillance-oriented information articles lend themselves naturally to topic-based stratification where strata are defined based on the subject of the story. An objective is to ensure individual articles within a stratum are diversified so that pairwise-similarity is low.

It is also worth noting that the surveillance aspect of news can often be looked at in the context of the whole set as well. For instance, if we have already included an article on a curfew lockdown in the list, the algorithm should not recommend another one on the topic even though it is about surveillance. So the goal is to make sure the surveillance articles cover different articles that the user needs to know but they should not be repetitive. This aggregate view of surveillance is a research challenge and we discuss it further in our future work section.

\subsection{News serendipity}

The second main stratum goes beyond information needs and brings the user joy, surprise, diversion, relaxation, etc. These traits are related to the concept of \emph{serendipity} in RS \cite{kotkov2016survey} although with some unique characteristics. According to the dictionary, serendipity is ``the faculty of making
fortunate discoveries by accident.''\footnote{https://www.thefreedictionary.com/serendipity} Gemmis et al. \cite{de2015investigation} defines \emph{serendipity} in recommender systems as those suggestions that are both \textit{attractive} (or joyful) and \textit{unexpected}. Our idea is to intersperse articles from this strata among the surveillance articles to provide these \textit{fortunate discoveries}. 

\begin{figure}
    \centering
    \includegraphics[width=2.8in]{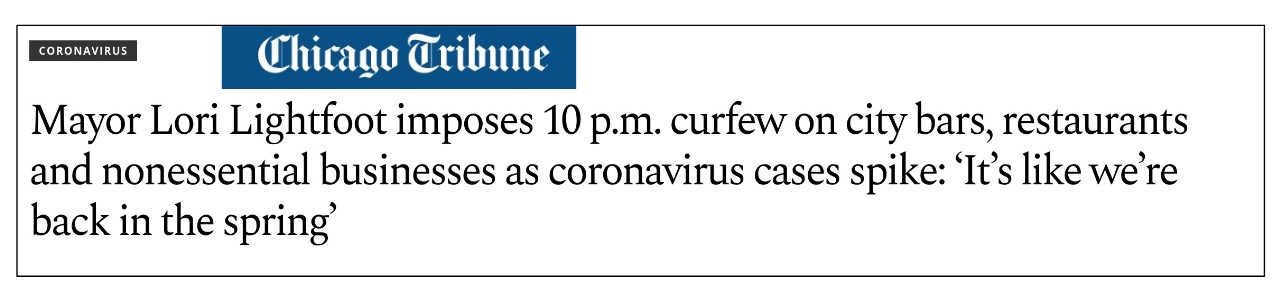}
    \includegraphics[width=2.77in]{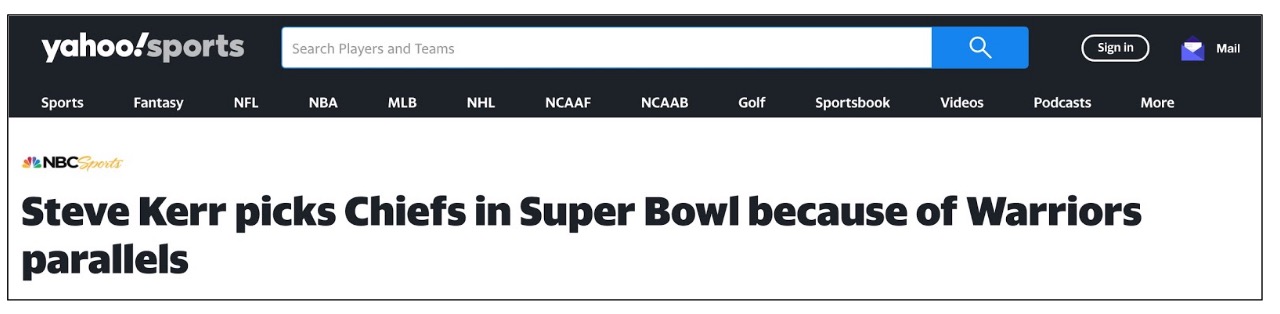}
    \includegraphics[width=2.8in]{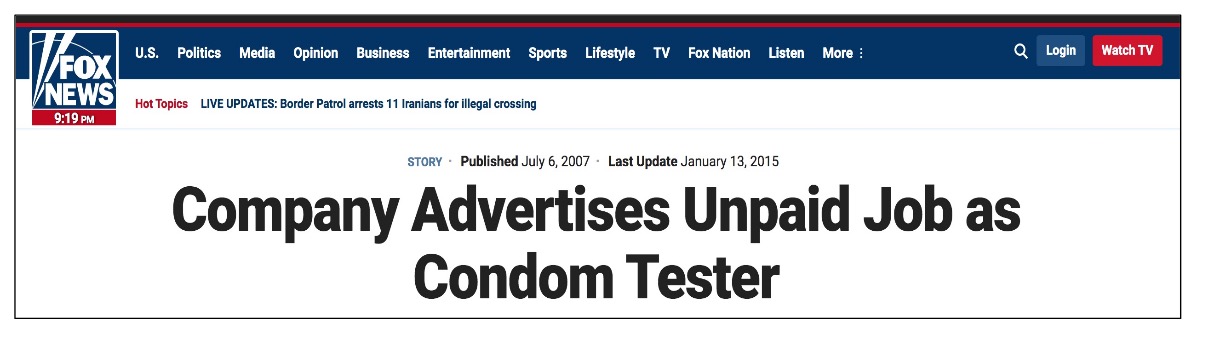}
    \caption{Examples of non-personalized surveillance (top), personalized surveillance (middle), and serendipitous (bottom) articles.}
    \label{fig:fig_examples}
\end{figure}

There are different ways that news can create serendipity for users. The Calder-Malthouse reader experience studies \cite{calder2002NPexp, Calder2009,mersey2010} attempted to measure how news media satisfy reader needs and they developed a typology \cite{Peck2011} of user needs. Many of their types fit under the news serendipity concept we have in mind, e.g.,

\begin{itemize}
\item \textbf{Talkers}. News often creates a social experience for users by giving them something to talk about with their family and friends. Thus, reading news helps them connect with others and maintain their relationships. News media can facilitate such interactions with talker stories \cite[p.\ 178]{Peck2011}. The bottom picture in Figure~\ref{fig:fig_examples} gives an example of a talker. One can imagine such stories being mentioned at the water cooler.
\item \textbf{Feel good}. Surveillance stories are, by their nature, often bad news and readers often feel helpless and depressed after reading a series of negative stories. While not all news can be good, readers appreciate articles that make them feel good. The story could, for example, have a happy ending (celebrity overcomes drug addiction and goes on to help others) or show that there are good people in the world. Several magazines have, as a core part of their concept, feel-good stories, e.g., \emph{Glamour} and \emph{Guideposts} \cite[p.\ 15]{Peck2011}.
    \item \textbf{Utilitarian}: give the user tips and advice, e.g., something fun to do this weekend, a recipe for happy hour drink, etc.
    \item \textbf{Images that wow, interactive visuals:} A beautiful photography slipped into a list of news articles could really bring joy to the user \cite[pp.\ 165--176]{Peck2011}. 
    \item \textbf{Entertainment and diversion}: e.g., a funny story or trivia quiz \cite[pp.\ 155-164]{Peck2011}.
    \item \textbf{Timeouts}: These are the articles that help the users escape from their daily stress by shifting their focus on a long, enjoyable story \cite[pp.\ 141-154]{Peck2011}. A related construct is \textbf{media transportation} \cite{wang2006media}, where users are mentally transported to a different place through the narrative. An example could be a suspenseful adventure story in the travel section.
\end{itemize}


Allocating some portion of the news list to the serendipitous articles that can bring joy and surprise to the users can give them a better experience so that they feel that the news is not just articles that scare them. See \cite[appendix]{Peck2011} for a complete list of experiences and prototypical quotes from user interviews. 



\section{Creating a \tabledhote{} news RS}

Creating \tabledhote{} goes beyond an RS algorithm and is a multi-objective process. On the one hand, the articles need to be classified into surveillance and serendipity. First and foremost, we need a certain process that can determine whether a certain article is covering an information need (whether critical or just a casual interest) or it is more a joyful, surprising article. Complicating matters further, the classification could vary over users, where an article that is serendipitous for one user is surveillance for another. 

\begin{figure}
    \centering
    \includegraphics[width=3in]{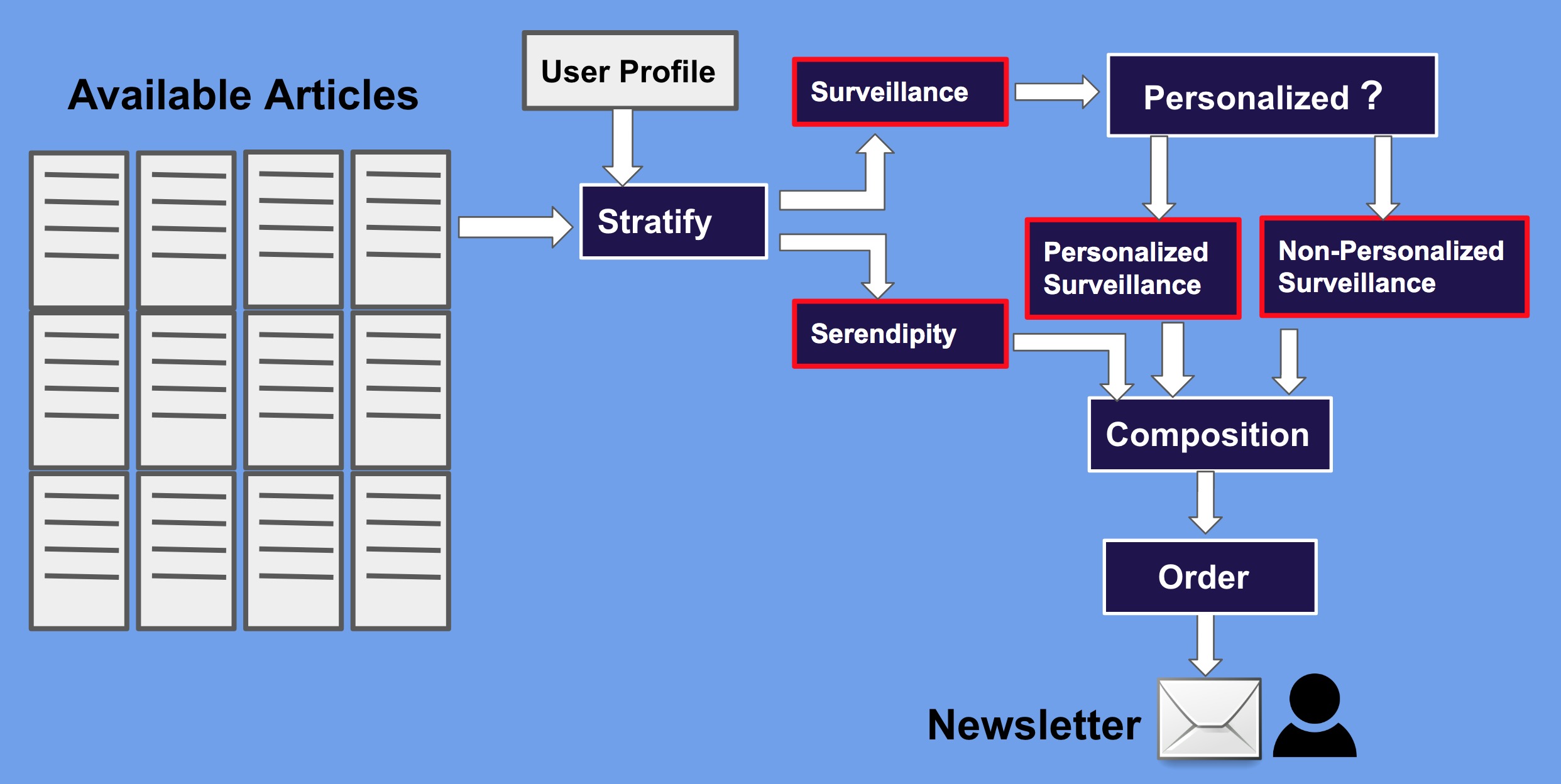}
    \caption{The mechanism of creating \tabledhote}
    \label{mechansim}
\end{figure}

This step can be done either manually by a human editor or it could be learned via machine learning algorithms that were previously trained to do such task using previous news articles. In addition, the surveillance articles need to be further categorized into personalized and non-personalized. Next, once the system classifies each article, another component should decide what proportion of the newsletter should be allocated to each of these different types. This part could be either defined by the \textit{purpose \& mission} of the organization that is sending out the newsletters or it could be algorithmically defined to optimize some objective. Finally, the order in which each article within the \tabledhote{} should be consumed by each user needs to be defined. Figure \ref{mechansim} summarizes the mechanism of the \tabledhote{} approach.

\subsection{Sequence of news articles in \tabledhote{} }
The \tabledhote{} approach extends the diversity of the recommendations from a set property (i.e., order does not matter) to a ranked-list property where the diversification needs not only to cover different aspects but also in a specific order. Similar to the chef example where the items within a prepared meal need to be served in a certain order to give the guest the best dining experience, the order of how the user reads the news articles within a \tabledhote{} can also be important for a good reading experience. For instance, should the list start with a happy news, a fun article, or a trending, important, article? Where should we put the serendipitous articles in the list? In the beginning, somewhere in the middle, or keep it for the end? In addition, some news articles may inherently need a certain level of audience preparation in order for them to be accepted and trusted by the audience. For instance, imagine an article that contradicts what partisan news organizations are saying about a topic. Should the article be first in the list, or should it be in a later position after showing other articles to build the user's trust? These are important questions that have not attracted much attention in the news domain. The importance of the sequence in RS has already been studied in other domains such as in music \cite{wang2018sequence} and requires more attention in the news domain as well. 
\section{Future \tabledhote{} research agenda}
The \tabledhote{} approach opens up many novel challenges and opportunities for further research:

\begin{itemize}

    \item 
    \textbf{Stratification:} A natural question is how to define the types of articles? For example, is the article a fun, joyful story, or does it convey critical, life-saving information. While a human may easily be able to distinguish between such articles, it is not a trivial task for an automated AI algorithm to do that. Techniques such as sentiment analysis could be applied to detect humor or ``feel-good'' \cite{barbieri2014automatic} but those also need a previous human supervision to learn some patterns such as important features that could indicate humor. In addition, serendipitous articles are not only about humor. \textit{Talker} articles, for instance, could be even about non-humorous topics that can help readers have something to talk about with their friends or family. Therefore, this aspect of \tabledhote{} has some challenges that are not yet fully explored.

    \item 
    \textbf{Composition}: Another interesting topic is how to determine the proportion of articles to be selected from each stratum. Should the organization allocate a fixed ratio of different types of articles and apply that to all users or it could be defined algorithmically for each user. For instance, should we keep talker articles in someone's newsletter even though s/he has not yet clicked on any during the past couple of interactions? It is worth noting that the fixed ratio set by the organization can be also defined via a data-centric approach where the historical interactions of the users are investigated and the ratios that could bring the highest amount of engagement could be set. 
    
    \item \textbf{Presentation}: As with a great meal, the presentation of the \tabledhote{} may be as important as the story selection. Should there be snippets summarizing the story, images or advertisements? If so, should they be personalized? If so, how to automate the personalization? Should there be a consistent tone or voice? For example, a no-nonsense, financial publication would want a different look and feel than a neighborly, small-town newspaper, which could imply different strata definitions and proportions.
    \item 
    \textbf{Effectiveness}: We need to know if the \tabledhote{} approach outperforms existing RS. This likely requires testing on real users and measuring the right outcomes. In the case of a news organization or an aggregator like Apple News that relies on subscription revenue, one important downstream goal is subscriber retention, although there may also be a another, possibly competing, objective to generate advertising revenues. What are leading indicators or these long-term goals? Answers likely begin with using the \tabledhote{} RS regularly, e.g., opening and spending time with newsletters/e-editions or visiting the site often. There are also  opportunities to develop new metrics for the specific types of diversity and serendipity implied by \tabledhote. From a social-good viewpoint, an important outcome is whether \tabledhote{} RS readers become better informed citizens.
\end{itemize}

\section{Conclusion}
In this paper, we introduced the concept of \tabledhote{} for news RS where the goal is to prepare a set of articles that together can fulfill the user's needs for information and joy. We discussed different types of content that can go into a \tabledhote: 1) \textit{surveillance} which are the articles that are either critical and need to be read by all or at least some users, and 2) \textit{serendipity} which are the articles that may not necessarily be related to the humans desire for information but rather to bring them joy and surprise. The organization then needs to decide how the articles can go into any of these strata and to what proportion and in what order they should be sent to the users. We believe the \tabledhote{} approach can be an effective way to humanize the news audience by providing them more than just news articles that may inform people about some important events but rather to prepare a meal for them that is fully customized and ready to serve which covers different aspects of their needs.


\bibliographystyle{ACM-Reference-Format}
\balance
\bibliography{main}
\end{document}